\documentclass{jfm}

\usepackage{graphicx}
\usepackage{newtxtext}
\usepackage{newtxmath}
\usepackage{natbib}
\usepackage{hyperref}
\hypersetup{
    colorlinks = true,
    urlcolor   = blue,
    citecolor  = black,
}

\newcommand{\RomanNumeralCaps}[1]
\linenumbers

\usepackage[utf8]{inputenc}         
\usepackage[subrefformat=parens,labelformat=parens]{subfig}
\usepackage{caption} 
\usepackage[utf8]{inputenc}
\usepackage[export]{adjustbox}  
\usepackage{bm}     
\usepackage{amsmath}     

\title{The interplay of vortex rings and streaks in the near field of turbulent jets}

\author{Jahnavi Kantharaju\aff{1}
  \corresp{\email{jahnavi.kantharaju@gmail.com}},
  Benjamin Leclaire\aff{1}
  \and
  Laurent Jacquin\aff{2}
  }

\affiliation{\aff{1} DAAA, ONERA, Institut Polytechnique de Paris, 92190 Meudon, France \\
\aff{2} ONERA, F-91123 Palaiseau, France}

\begin{document}
\maketitle

\begin{abstract}

In light of recent renewed interest in streaks in turbulent jets
, the current work explores their coexistence with vortex rings in the near field of turbulent round jets. A Reynolds number, $Re=1.5 \times 10^5$ jet is studied at two diameters downstream of the nozzle, using high-speed stereo particle image velocimetry. The spectra of individual velocity components reveal radially localized signatures of the large scale structures.  The radial shapes of the spectral proper orthogonal decomposition modes corresponding to the signatures, confirm the presence of high speed streaks in the outer edge of the jet at low Strouhal number, $St \rightarrow 0$. The vortex rings and streamwise vortices are found to occur in the shear layer at $St \approx 0.49$, where they convect together as a system and feed the streaks. The role of vortex rings in the existence of streaks is then studied by strengthening the rings through axisymmetric excitation. The streaks are observed to persist, retaining their shapes and show only slight changes in their energies.

\end{abstract}


\section{Introduction} \label{sect:intro}

A rich dynamics exists in the near field of round jets due to the interactions between different large scale structures. One of the important structures evidenced in the visualizations of jets at low Reynolds numbers, $Re$ is the vortex rings, which are a result of the roll-up of the shear layer subjected to Kelvin-Helmholtz instability. The other is the secondary structure in the form of streamwise vortices, that are formed as a result of a secondary instability of either the vortex rings \citep{Pierrehumbert1982} or the vorticity in the braid region \citep{corcos1984mixing2}. However, at higher $Re$, the incoming boundary layer becomes turbulent and the presence of small-scale, less organized turbulence leads to unsteady and irregular formation of the above discussed vortices. Several works such as \cite{Citriniti2000,tinney2008low} have extracted the large scale structures in turbulent jets using spectral proper orthogonal decomposition, SPOD (see e.g. \cite{towne2018spectral}) based on turbulent kinetic energy.

The existence of streaks, the streamwise elongated structures, extensively studied in wall-bounded flows \citep{kline1967structure}, has only recently been explored in turbulent round jets \citep{nogueira2019large,pickering2020}. Theoretical works such as \cite{jimenez2017, marant2018influence,wang2020effect} have found that transient growth analysis of streamwise invariant disturbances yields streamwise vortices as the optimal initial condition and streaks as the optimally amplified disturbances. 
The analysis of \cite{marant2018influence} showed that streaks can modify the characteristics of KH instability. These two studies have motivated further investigation of streaks in turbulent jets, especially in the direction of devising control strategies to by-pass KH instability that contributes greatly to jet noise \citep{jordan2013wave}.

In the earlier studies, \cite{Citriniti2000,tinney2008low} observed entrainment and ejection of fluid into and out of the jet core, in their reconstructed velocity field from dominant SPOD modes. However, the resulting structures were not particularly associated with streaks in these works. The ejected fluid was seen to remain in the measurement plane for long times, suggesting their almost steady nature. While it can be expected that the presence of an array of oppositely rotating streamwise vortices can result in strong radial ejections, the fact that the ejected fluid can persist in the outer edge of the shear layer for long times is quite fascinating. It is also interesting to note that unlike in turbulent boundary layers, the discovery of streaks comes rather recently compared to more than half a decade of research on understanding streamwise vortices. \cite{Monkewitz1991} reported intense side-jets rather than streaks-like structures in highly excited jets due to both vortex rings and streamwise vortices. Hence, the existence of streaks in turbulent jets may not be trivial.

Comparing with streaks found in turbulent boundary layers, the experiments of \cite{nogueira2019large} showed the relevance of large-scale streaks in jet dynamics. \cite{pickering2020} studied different amplification mechanisms such as Kelvin-Helmholtz, lift-up and Orr mechanism that are active in the near field of turbulent jets. Through global resolvent analysis and SPOD, \cite{pickering2020} showed that lift-up mechanism is dominant for very low frequencies, $St$ and non-zero $m$. The optimal forcing structures that result in the observed streaks were in the form of streamwise vortices at $St \rightarrow 0$ near the nozzle exit. The question then remains about the origin of these streamwise vortices at $St \rightarrow 0$ and the role of streamwise vortices that have been reported earlier to be convecting with the vortex rings \citep{Liepmann1992,Citriniti2000, kantharaju2020}, in the existence of the observed streaks. The presence of vortex rings along with streamwise vortices could set them apart from their counterparts in turbulent boundary layers, the exploration of which, is one of the aims of the current work.

Here, we seek more insights into the co-existence of streaks with the other structures in turbulent round jets with tripped boundary layers. The novelty lies in placing the steady streaks in the existing picture of the near-field of jets consisting of convecting system of vortex rings and streamwise vortices, and studying the effect of vortex rings on streaks through axisymmetric excitation. We show that the streamwise vortices occurring at the same frequency as that of the rings, feed to these steady streaks.

We begin with a brief note on the experimental database and SPOD formulation in section \ref{sect:methods}. The coherent structures in an unexcited round jet at $Re=1.5 \times 10^5$ are studied in section \ref{sect:unforced}. Section \ref{sect:excited} explores the effect of axisymmetric excitation on the streaks.

 \section{Method}\label{sect:methods}
 \begin{figure}
\centering

  \includegraphics[scale=0.15]{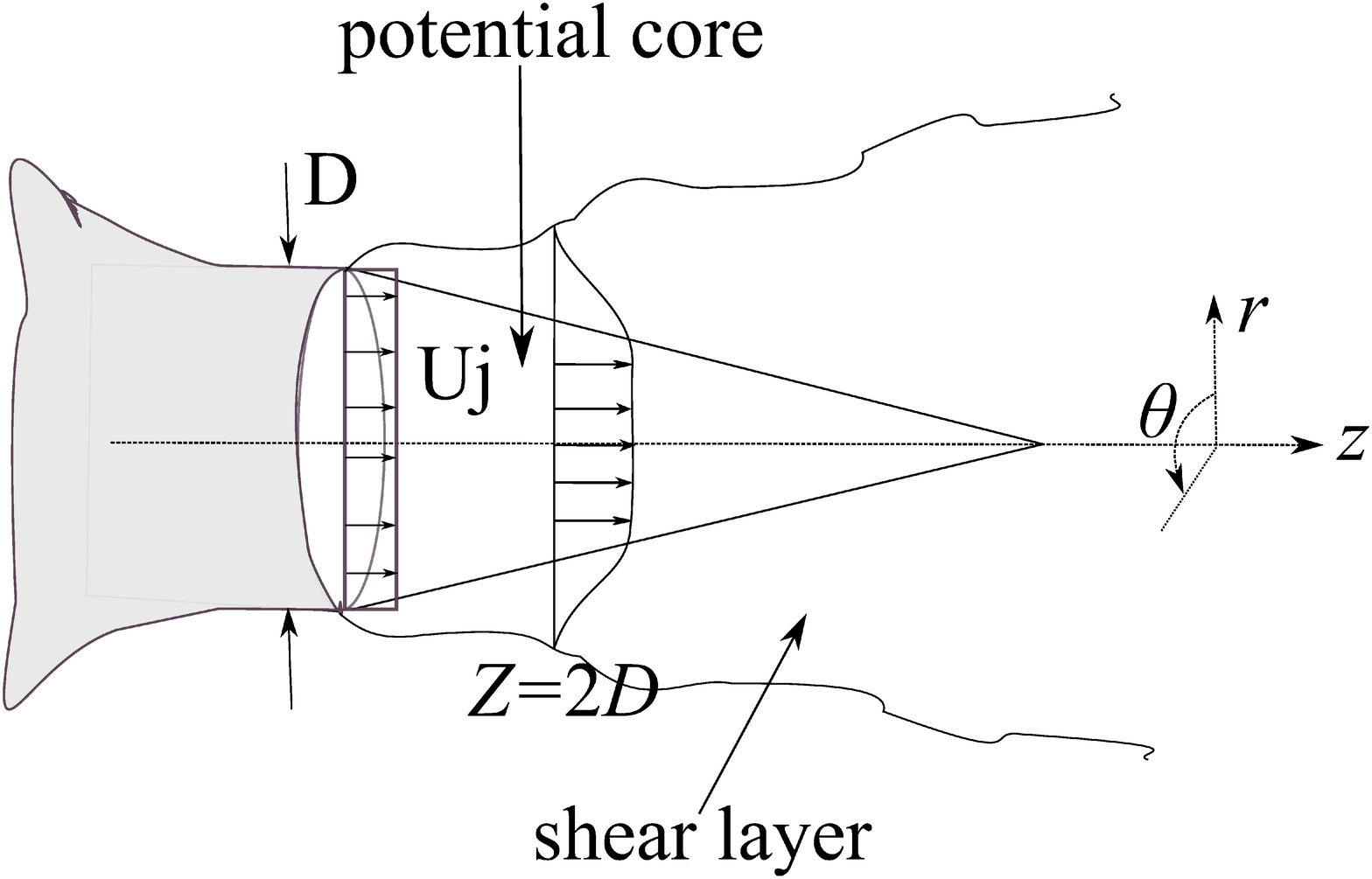}

\caption{Definition of the jet flow studied}
\label{fig:jetDef}
\end{figure}

The experimental database of a $Re=1.5 \times 10^5$ round jet with tripped boundary layer, described in \cite{kantharaju2020} is analyzed in the current work. High-speed stereo particle image velocimetry was performed in a cross-sectional plane at two diameters downstream of the nozzle exit, $Z=2D$. The acquisition frequency was 2.5 kHz and the spatial resolution of the vector field was $0.018D$.  

Axisymmetric excitation was achieved by means of a loudspeaker mounted on top of the settling chamber, that provided axisymmetric forcing at the nozzle exit. The loudspeaker was driven by a pure sine wave whose amplitude can be varied through the voltage supplied to the loudspeaker. The excitation level was characterized in terms of the root-mean-square of the axial velocity fluctuations at the nozzle exit centerline.

Spectral proper orthogonal decomposition (SPOD) was then used to extract the most energetic structures in the flow. An algorithm similar to \cite{towne2018spectral} was implemented to compute the SPOD basis, with 256 blocks of 512 samples each with an overlap of 50 \%, yielding a frequency resolution of 4.88 Hz ($St \approx 0.03$). The SPOD formulation and implementation can be further found in \cite{kantharaju2020}. 

In the following sections, we represent non-dimensional quantities by lower-case ($r,\theta,z$) symbols in the cylindrical coordinate system. Lengths and velocities are presented in non-dimensional forms, scaled with the nozzle exit diameter ($D$) and centerline average jet velocity ($U_j$), respectively. Non-dimensional time is expressed by $t$ and frequency by Strouhal number, $St=F D/U_j$, with $St_e$ denoting the excitation Strouhal number. We denote the components of the mean velocity vector by ($u_r,u_\theta,u_z$), and the fluctuations by a prime ($u'_r,u'_\theta,u'_z$).

 
\section{Streaks in the presence of vortex rings and streamwise vortices}\label{sect:unforced}

\subsection{Signature of the coherent structures in the energy spectra}

\begin{figure}
\centering
\subfloat[]{%
  \includegraphics[scale=0.65]{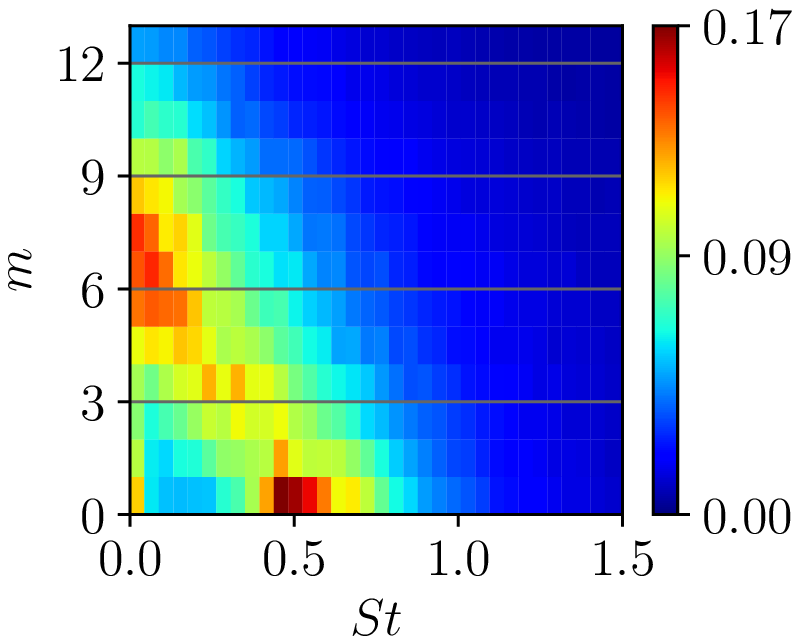}
  \label{fig:TE_1_nf}
}
\subfloat[]{%
  \includegraphics[scale=0.65]{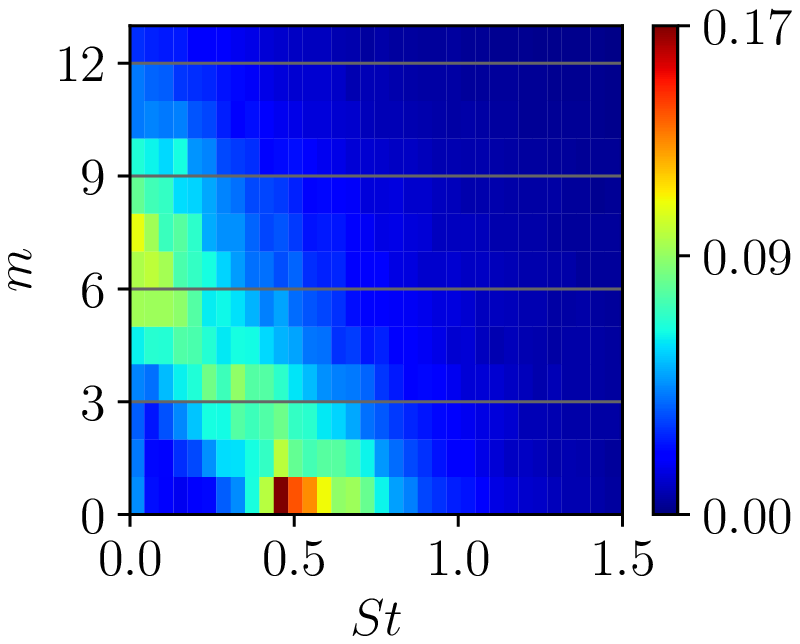}
  \label{fig:1_nf_spodEnergies}
}

\caption{Distribution of the turbulent kinetic energy at $z=2$ across different frequency $St$ and azimuthal mode $m$ a) total energy b) in the $n=1$ SPOD mode
}
\end{figure}

We begin by looking at the distribution of the total turbulent kinetic energy, 
in the cross-sectional plane at $z=2$, as a function of $St$ and $m$, and is shown in figure \ref{fig:TE_1_nf}. 
The energy is seen to be concentrated in two regions, $viz.$ 1) $m=0$ around $St=0.49$ which is a signature of the vortex rings \citep{Citriniti2000,kantharaju2020} and 2) around $m=6$, $St \rightarrow 0$ that was recently reported to be representing high-speed streaks \citep{nogueira2019large, pickering2020}. There appear to be some spurious values at $m=0$, $St=0.03$, the possible source of this is discussed below.

\begin{figure}
\subfloat[Total energy]{%
  \includegraphics[scale=0.65,valign=t]{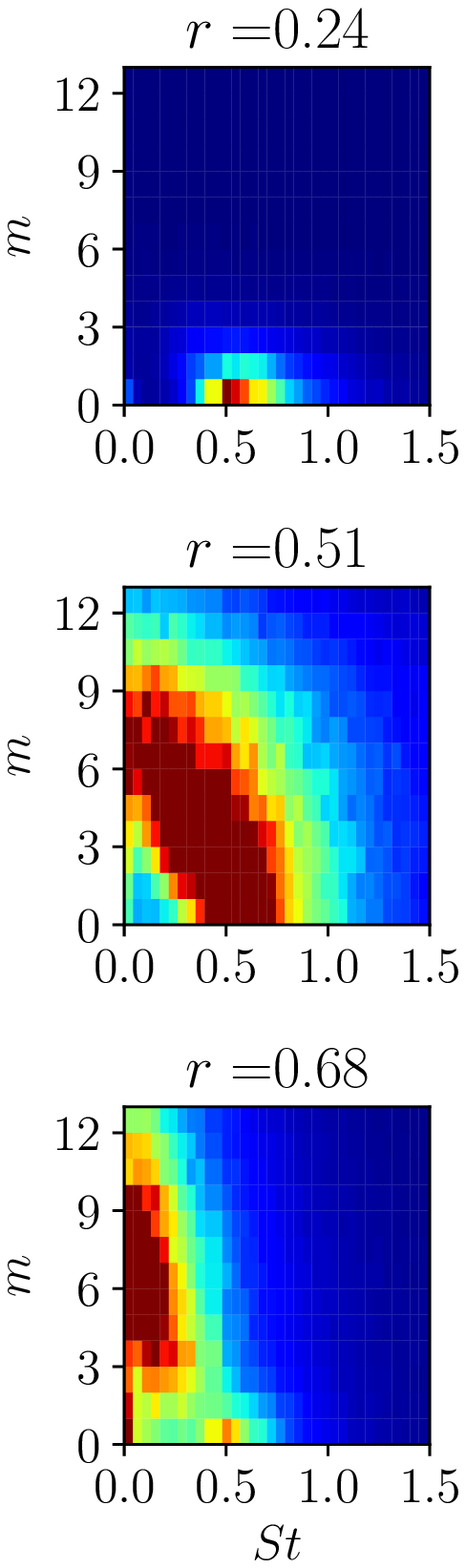}
  \label{fig:TE_1_nf_radial}
}
\subfloat[$u'_z$]{%
  \includegraphics[scale=0.65,valign=t]{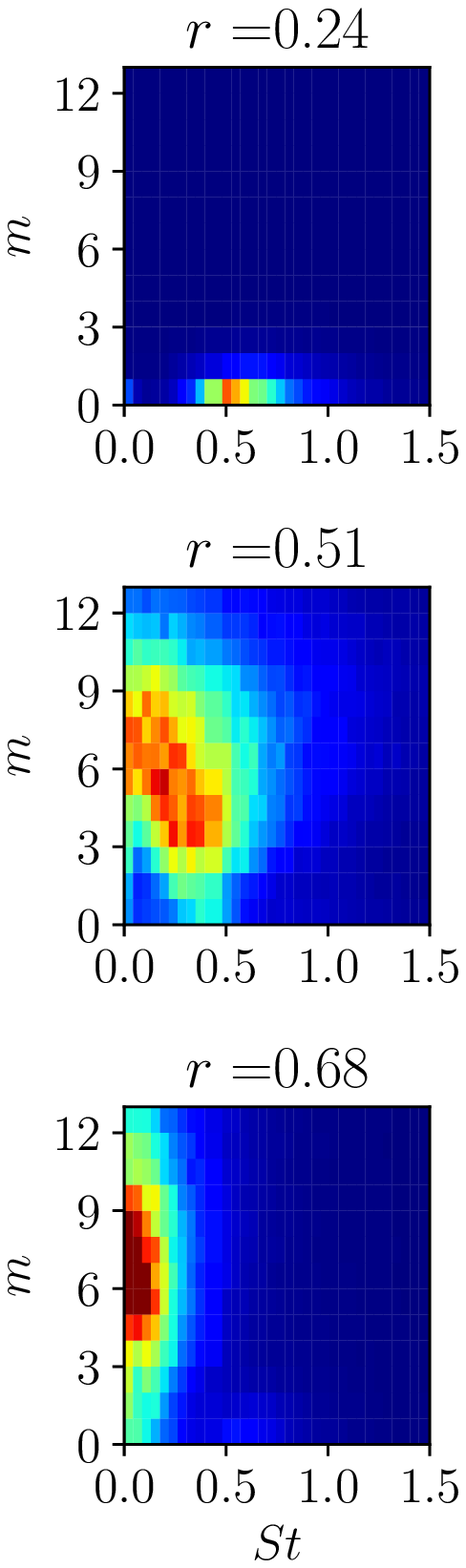}

}
\subfloat[$u'_r$]{%
  \includegraphics[scale=0.65,valign=t]{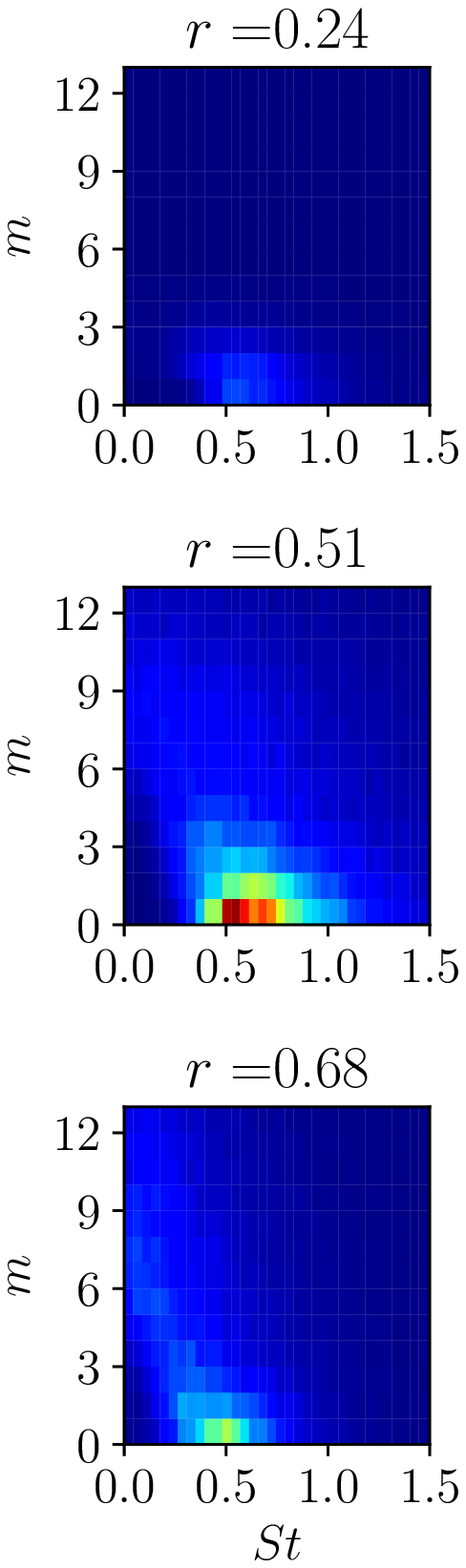}

}
\subfloat[$u'_\theta$]{%
  \includegraphics[scale=0.65,valign=t]{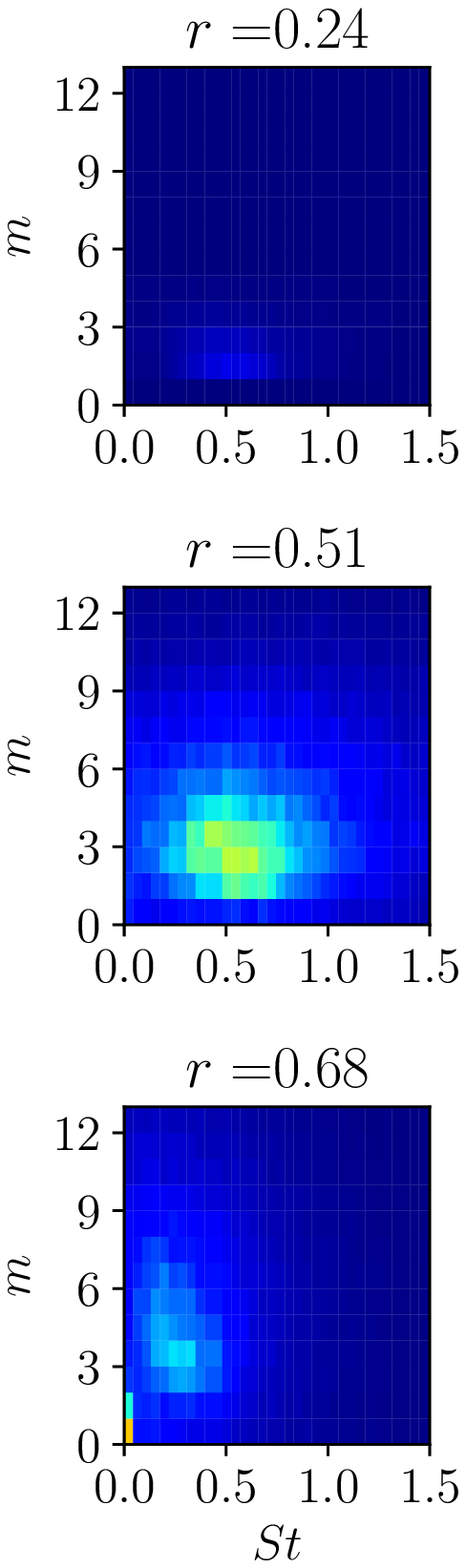}
  \label{fig:uth_tke}
  
}
\subfloat{%
\raisebox{-55ex}{
  \includegraphics[scale=0.65]{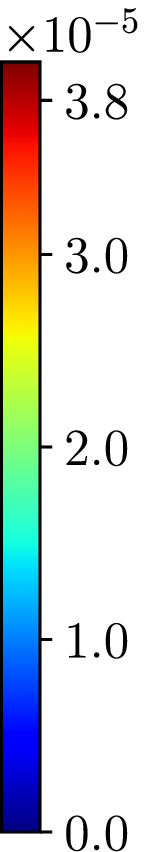}}

}
\caption{Total and individual turbulent kinetic energy at various radial locations as indicated above the plots.}
\label{fig:TE_1_nf_indCom}
\end{figure}

The above energy distribution can then be viewed at different radial locations 
as given in figure \ref{fig:TE_1_nf_radial}. As we move away from the jet centerline, the energy concentration starting around $m=0$, $St=0.49$ in the core, gradually shifts towards higher $m$ and lower $St$. In the shear layer at $r=0.51$, as expected, a wide spectrum of various frequencies and azimuthal modes is evident. Thus a spatial localization of the turbulent kinetic energy along the radius is present in the $St-m$ space.

The coherent structures discussed in the introduction section \ref{sect:intro}, have characteristic velocity components. Hence, the contribution from individual velocity components to the total energy could provide further insights on the structure of the energetic modes and is included in the rest of the figure \ref{fig:TE_1_nf_indCom}. The axial velocity spectra contain most of the energy observed in figure \ref{fig:TE_1_nf_radial}. In the outer edge of the shear layer, the signature of the streaks can be found to occur at very low $St \rightarrow 0$, i.e. the streaks are almost steady with azimuthal wavenumbers around $m=6$. The in-plane velocity components, $u'_r,u'_\theta$, on the other hand, have energies around $St=0.49$ and a low dimensional nature in terms of $m$ compared to $u'_z$, as observed by \cite{tinney2008low}. Note that from figure \ref{fig:uth_tke}, the spurious energy seen in figure \ref{fig:TE_1_nf} at $m=0$ and $St\rightarrow0$, is present in the $u_\theta$ component in the outer edge of the shear layer. It is suspected that a geometrical perturbation due to the room layout could be responsible for breaking the azimuthal symmetry, resulting in affecting the flow in the outer edge. 

\subsection{Deducing radial shapes of the energetic modes using SPOD}

Having identified the energetic regions in the spectra, we resort to SPOD in order to deduce the spatial structure of the corresponding modes. The first two SPOD modes capture upto 55\% and 19\% of the total turbulent kinetic energy, respectively. Figure \ref{fig:1_nf_spodEnergies} presents the energy spectra
of the $n=1$ 
SPOD mode. The overall trend of the total energy in figure \ref{fig:TE_1_nf} can be seen in the $n=1$ SPOD mode, thus the large scale dynamics in the current flow can be approximately described by just considering the $n=1$ mode. The radial distribution and contribution from individual velocity components for the $n=1$ SPOD mode followed trends similar to those of the total energy (figure \ref{fig:TE_1_nf_indCom}). 

\begin{figure}
\centering
\subfloat[]{%
  \includegraphics[scale=0.65]{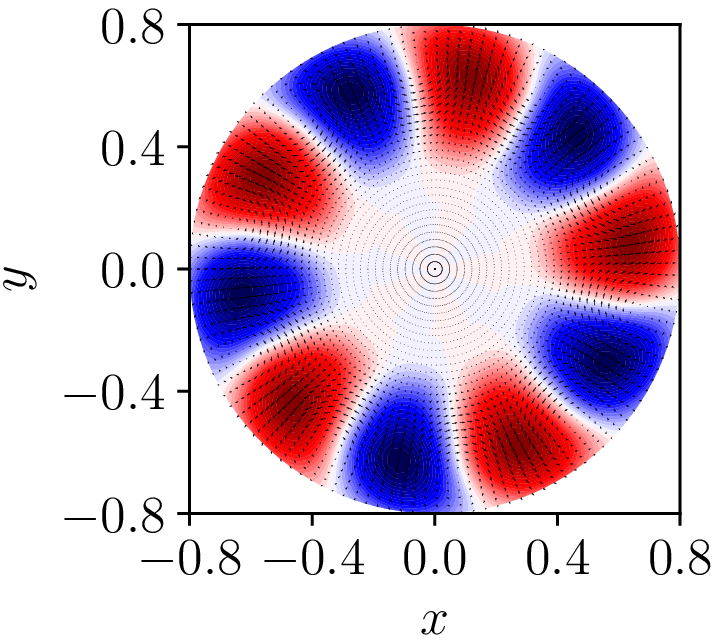}
  \label{fig:st0}
}
\subfloat[]{%
  \includegraphics[scale=0.65]{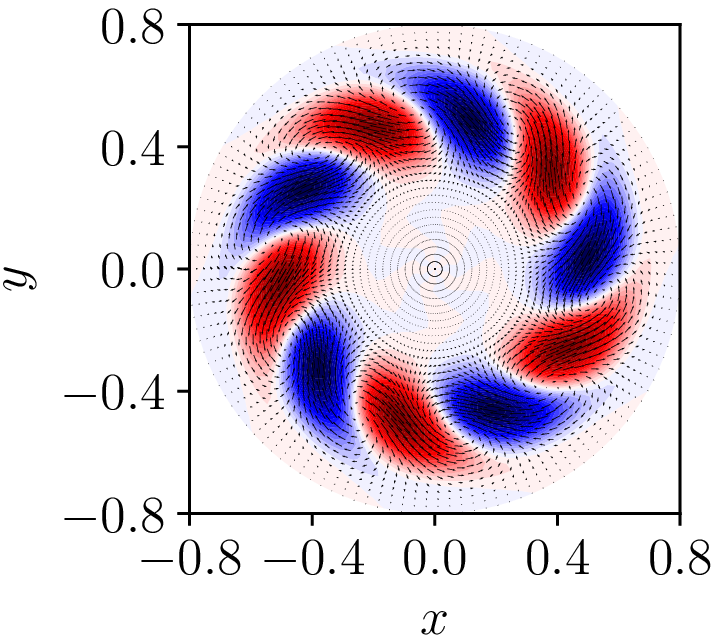}
  \label{fig:st0.5}
}
\vspace*{-0.05cm} 
\subfloat{%
\raisebox{3ex}{
  \includegraphics[scale=0.65]{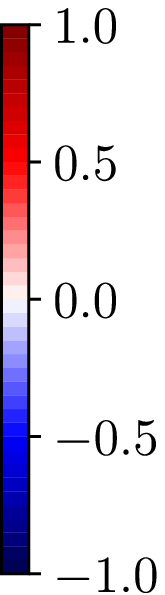}}

}
\caption{$n=1$ SPOD mode corresponding to $m=5$, a) $St=0.03$ b) $St=0.49$; $u'_z$ are shown by contours and $u'_r$, $u'_\theta$ as vectors. 
}
\label{fig:1_nf_eig}
\end{figure}

Figure \ref{fig:1_nf_eig} shows the shape of the modes in the energetic regions identified in the previous section, viz. $m=5$ at $St=0.03$ and $St=0.49$. The contours indicate $u'_z$ and the vectors represent the in-plane velocity components, ($u'_r$, $u'_\theta$). Although the shapes of both these modes represent streamwise vortices and the associated alternate high and low axial velocity patches, the energy in the individual velocity components (figure \ref{fig:TE_1_nf_indCom}) indicates that the streamwise vortices associated with $St=0.03$ have significantly lower energy. 

The mode at $St=0.03$ in figure \ref{fig:st0} matches with the one reported in \cite{nogueira2019large, pickering2020}, corresponding to streaks. Whereas the mode in figure \ref{fig:st0.5} is in the form of streamwise vortices occurring around the same $St=0.49$ as that of the most energetic $m=0$ mode, i.e. the streamwise vortices convect with the vortex rings as previously observed in \cite{Liepmann1992,Citriniti2000,kantharaju2020}. These streamwise vortices are located in the shear layer and appear to be tilted in the azimuthal direction. The axial velocity contours show that the net induced velocity resulting from this arrangement of counter-rotating pair of streamwise vortices, lifts up high-momentum fluid from the jet core and draws in low-momentum fluid from the outside.

In the previous studies \citep{nogueira2019large,pickering2020}, the streamwise vortices associated with higher $St$ were not studied. However in the current study, these vortices (at $St=0.49$) are seen to have significant energy and play a role in the existence of streaks. The observed difference in the frequencies of the streamwise vortices at $St=0.49$ and the streaks at $St\rightarrow0$ in the current work, could stem from a scenario such as this: the high momentum fluid gets advected outwards by the resulting induced velocity of the convecting streamwise vortices. Owing to the low mean velocity in the outer edge of the shear layer, the advected fluid tends to reside in this region for longer times appearing as high-speed streaks and hence have signatures at $St \rightarrow 0$. The global resolvant analysis of \cite{pickering2020} showed that streaks are formed as a result of forcing from the streamwise vorticity near the nozzle exit. While this is one of the possibilities, the current analysis shows another scenario in which the streamwise vortices, formed from secondary instabilities as discussed in the introduction section \ref{sect:intro} and convect with the vortex rings, could contribute to the existence of these streaks, as was also hypothesized by \cite{Citriniti2000}. 

\subsection{Cross-correlation}

The association of streaks with the streamwise vortices at a given cross-sectional plane, i.e. at $z=2$, can be found by considering the correlation of $u'_z$ and $\omega'_z$ with time lag, $t'=0$ as:

\begin{equation} \label{eqn:vortCorr}
    C_{\omega_zu_z}(r,r',\theta',t')=\frac{<\omega'_z(r,\theta,t)u'_z(r',\theta+\theta',t+t')>_\theta}{<{\omega'_z}^2(r,\theta,t)>^{1/2}_\theta <{u'_z}^2(r',\theta,t)>^{1/2}_\theta}
\end{equation}

\noindent where $< >_\theta$ denotes average over both data blocks and $\theta$, as the flow has statistical stationarity and azimuthal homogeneity. 

\begin{figure}
\centering
\subfloat[$r=0.68$]{%
  \includegraphics[scale=1]{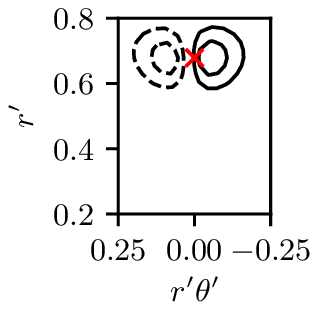}
  \label{st0}
}
\subfloat[$r=0.51$]{%
  \includegraphics[scale=1]{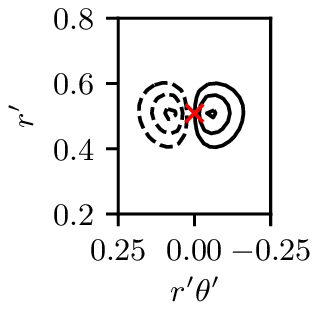}
  \label{st0}
}
\vspace*{-0.05cm} 
\subfloat[$r=0.38$]{%
  \includegraphics[scale=1]{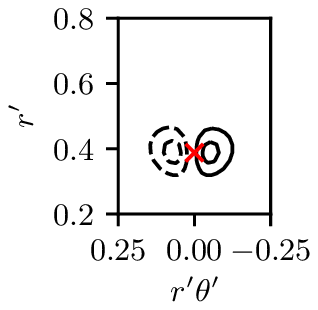}
  \label{st0}
}

\caption{$C_{\omega_zu_z}(r,r',\theta',t')$ for $t'=0$ and $r=0.39, 0.51, 0.68$; the probing point is marked by a red cross 'x'; the contour levels are (-0.3,-0.2,-0.1,0.1,0.2,0.3) starting with the highest at the center of the lobes on each side and decreasing with the size of the contours; negative values are shown in dashed lines and positive values with solid lines.}
\label{fig:oz_uz_corr_1_nf}
\end{figure}

This is shown in figure \ref{fig:oz_uz_corr_1_nf} at three radial locations at which $\omega'_z$ is probed, i.e. at $r=0.4, 0.51,0.68$. The probing points are marked as red crosses. Opposite-signed correlation maps can be found on the two sides of the probing point with growing size as we move away from the jet centerline. These correlation maps are similar to the image gained by figure \ref{fig:1_nf_eig}; opposite-signed streamwise vortices located between alternate signed $u'_z$ contours. Note the shift in the alignment of the probing point and the center of the patches as $r$ increases. Also, the lack of left-right symmetry about the probing point could stem from the slight tilting of the streaks in the azimuthal direction as in figure \ref{fig:1_nf_eig}. These correlations suggest the presence of streamwise vorticity and the associated induced axial velocity fluctuations across the cross-sectional plane, with the strongest correlation present in the shear layer as seen from the spectra plots.

\subsection{Reconstruction of the flow-field with the most energetic modes}

With the energetic SPOD modes in the $St-m$ space,  reduced order dynamics is obtained in this section, to better understand and isolate the large scale structures described above. The velocity vector was reconstructed similar to \cite{Citriniti2000}, from $m=0-12$, $n=1$ SPOD modes that well captured the large scale features of the current flow.

\begin{figure}
    \centering
    \includegraphics[width=5in]{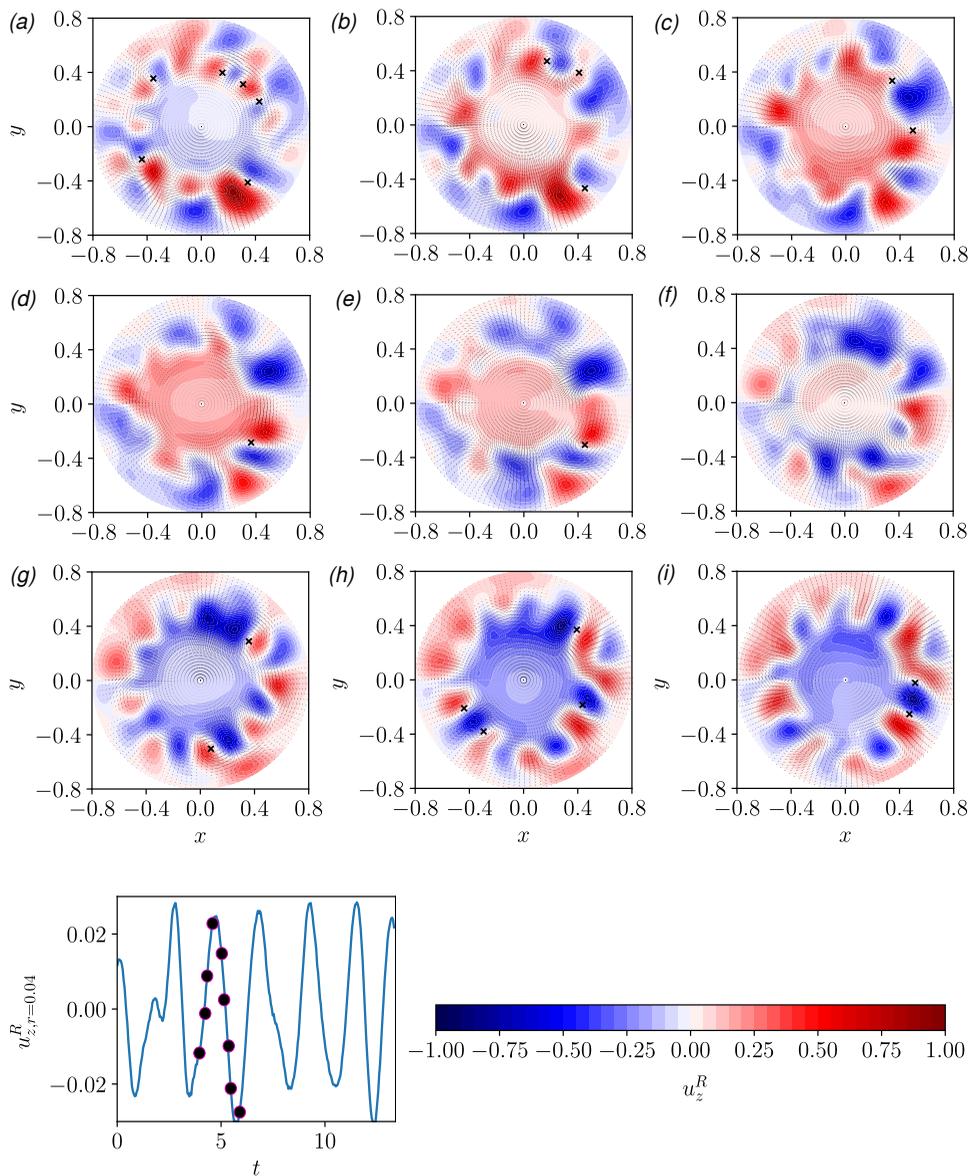}
    \caption{Contours of the reconstructed field $u'_z$ normalized by the maximum value over the time instances presented; reconstruction was done using the first $m=12$ azimuthal and $n=1$ SPOD mode at time instances $t=0.13$ apart; the vectors represent the in-plane velocity components and the possible locations of streamwise vortices are marked by a cross (\textbf{x}). }
    \label{fig:2Dcontours_reconstructed}
\end{figure}

Figure \ref{fig:2Dcontours_reconstructed} shows the contours of $u^R_z$ along with the in-plane velocity components as vectors, as a representation of an event around the passage of a vortex ring. This event is deduced from the time signal of $u^R_z$ in the jet core at $r=0.04$, 
included at the bottom of the figure, where the time instances of the contour plots are marked in the signal around a local maximum in $u'_z$ values, indicating the passage of a vortex ring. Localized swirling motions have been identified and marked by a cross (\textbf{x)}), that denote streamwise vortices. The high-low $u^R_z$ patches in subplot (a) in the shear layer can be seen to move outwards as the ring passes through. These patches then remain in the outer-edge for all time instances shown here and would appear as streaks when viewed along the axial direction. They increase and decrease in intensities as the ring passes through the plane. The appearance of the streamwise vortices, on the other hand, changes across the snapshots. \cite{Citriniti2000} presented similar contour plots of the reconstructed $u'_z$ from $n=1$, $m=0,3,4,5,6$ SPOD modes at $z=3$, where they identified possible locations of streamwise vortices that generate movement of high and low momentum fluid in and out of the jet core. The current analysis shows the definitive presence of streamwise vortices, endorsing the assertions of \cite{Citriniti2000}. 

From the analysis of this section, it can be asserted that the vortex rings, streamwise vortices and streaks coexist in the near field. The former two structures are found to be convecting through the measurement plane and feed the steady streaks located in the outer regions. The signatures of these three structures appear as a localization of the turbulent kinetic energy in the radial direction and at certain $St$.

\section{Effect of strengthening the vortex rings on the streaks}\label{sect:excited}

Next, we explore the co-existence of streaks and vortex rings. The vortex rings were strengthened through axisymmeteric excitation of the most energetic $m=0$ mode in the natural jet at $St_e=0.49$ (deduced in \cite{kantharaju2020}). The excitation amplitude was measured in terms of the fluctuation intensities at the nozzle exit, $u^{rms}_{z|z=0}$. In \cite{kantharaju2020}, it was estimated that the excitation at this $St_e=0.49$ results in an increase in the strength of the vortex rings relative to the streamwise vortices. Coherent streamwise vortices and rings were observed therein through three dimensional Taylor's reconstruction based on Taylor's hypothesis. 

\begin{figure}
\centering
\subfloat[$u'_z$]{%
  \includegraphics[scale=0.65,valign=t]{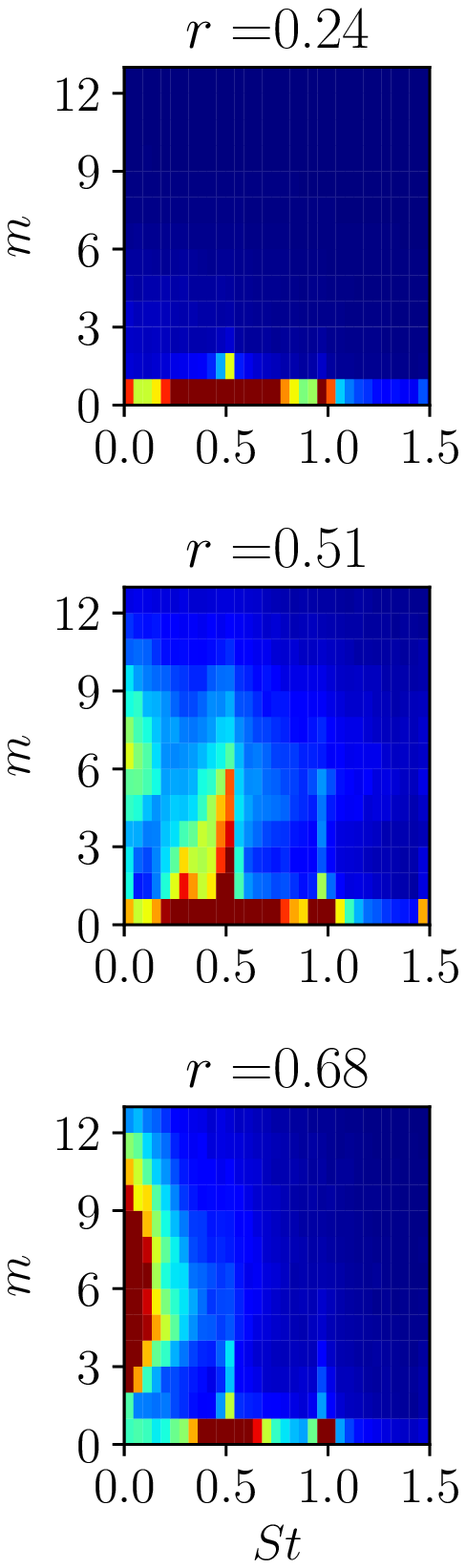}
}
\subfloat[$u'_r$]{%
  \includegraphics[scale=0.65,valign=t]{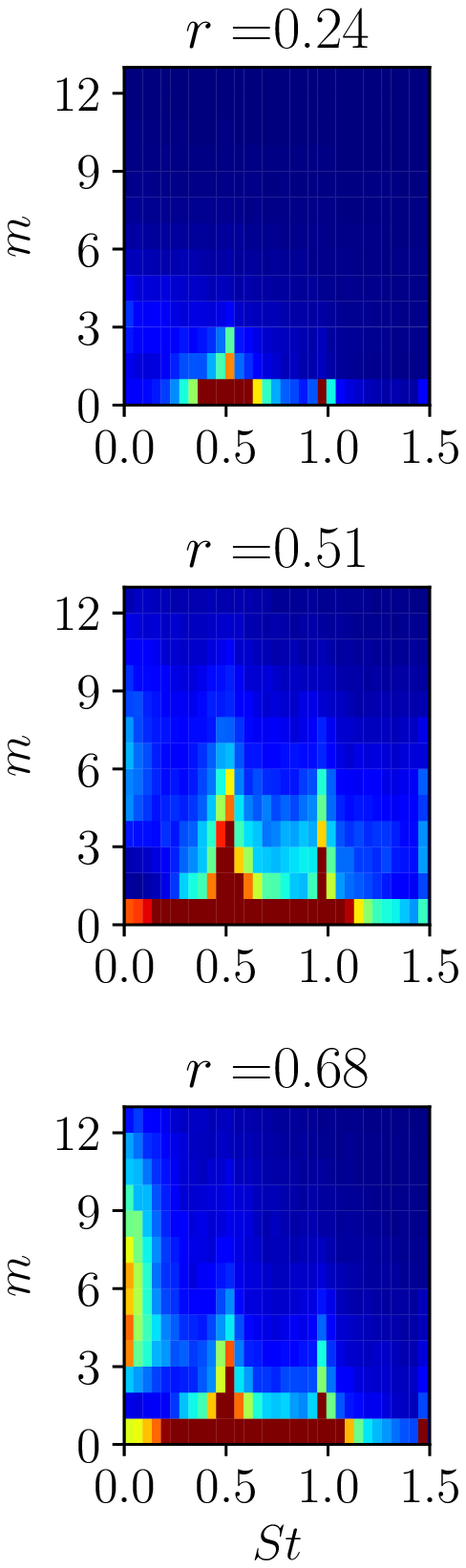}
}
\subfloat[$u'_\theta$]{%
  \includegraphics[scale=0.65,valign=t]{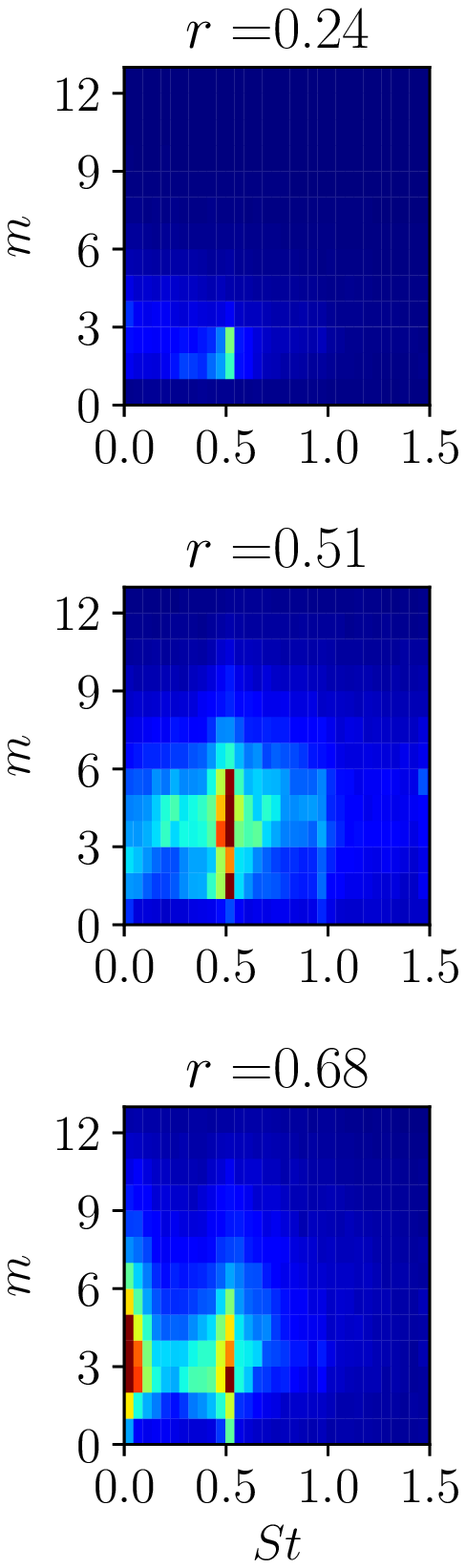}
  \label{fig:uth_tke}
  
}
\subfloat{%
\raisebox{-55ex}{
  \includegraphics[scale=0.65]{figsFeb/colorbar_spectra.eps}}
}
\caption{Turbulent kinetic energy for individual velocity components at various radial locations as indicated above the plots for }
\label{fig:excited_spectra}
\end{figure}

Figure \ref{fig:excited_spectra} contains the spectra of the individual velocity components at different radial locations for this excited case at $u^{rms}_{z|z=0}=2.6\%$ forcing. The $u'_z$ component has energy concentrated in the $m=0$ mode around the excited $St_e$ in the jet core. This is due to the strong $u'_z$ induced by the vortex rings. Also at $r=0.51$, some energy appears at $St\approx0.25$, a sub-harmonic of the forcing frequency, indicating the possibility of pairing of the vortex rings under these excitation conditions. 

In the outer edge of the shear layer, $u'_z$ is present at $St=0.03$, $m=[3,10]$ as well as at $St=0.49$, $m=0$. This implies that the fluctuations induced by the $m=0$ mode are seen upto $r=0.68$, denoting the increasing influence of the rings in the radial direction with excitation. An interesting observation from these spectra is the persistence of streaks even with the passage of the coherent rings, that tend to sweep the fluid ejected outwards by the streamwise vortices, through their induced velocities. There is a corresponding increase in the energies in $u'_r,u'_\theta$ at $St=0.49$ and its harmonics. Also in the outer edge of the shear layer, energy is seen to be present in the in-plane velocity components for $St=0.03$ and around $m=5$ which was not present in the spectra for the unforced jet. This suggests the presence of streamwise vortices in the outer-edge at low $St=0.03$ as well, similar to those reported by \cite{pickering2020}. Thus, in the excited jet in the current study, there are two sets of vortices observed, one at $St=0.49$ as seen for the unforced jet that convect with the vortex rings, and the other at $St=0.03$ which tend to remain steady. One possibility is that in the excited jet, the latter streamwise vortices could have broken from the first set of vortices and tend to reside in the outer-edge, or could be formed as a result of forcing from the streamwise vorticity near the nozzle exit as proposed by \cite{pickering2020}. This calls for further investigation into the origin of streamwise vortices in tripped jets. Previous studies such as \cite{Lin1984,Martin1991} have explored the formation of streamwise vorticity in plane shear layers and jets at lower Re.

\begin{figure}
\centering
\subfloat[]{%
  \includegraphics[scale=0.65]{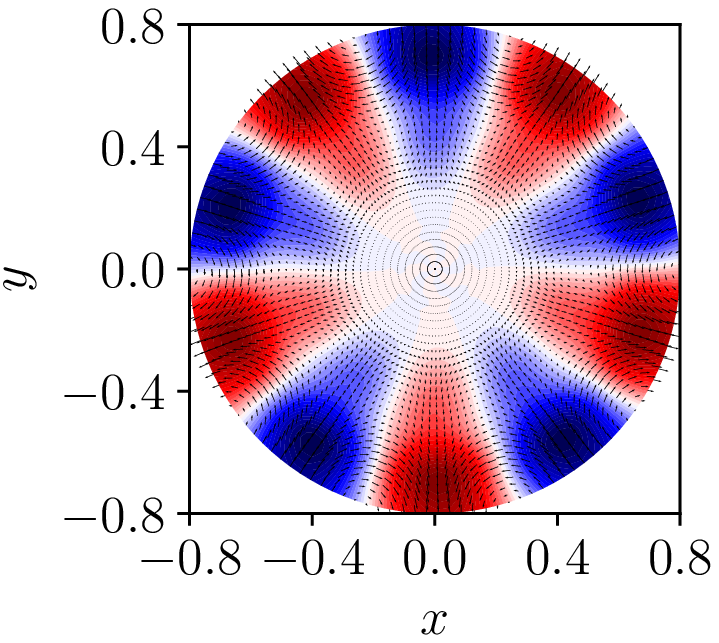}
  \label{fig:eigenFunc_excited_lowSt}
}
\vspace*{-0.05cm} 
\subfloat[]{%
  \includegraphics[scale=0.65]{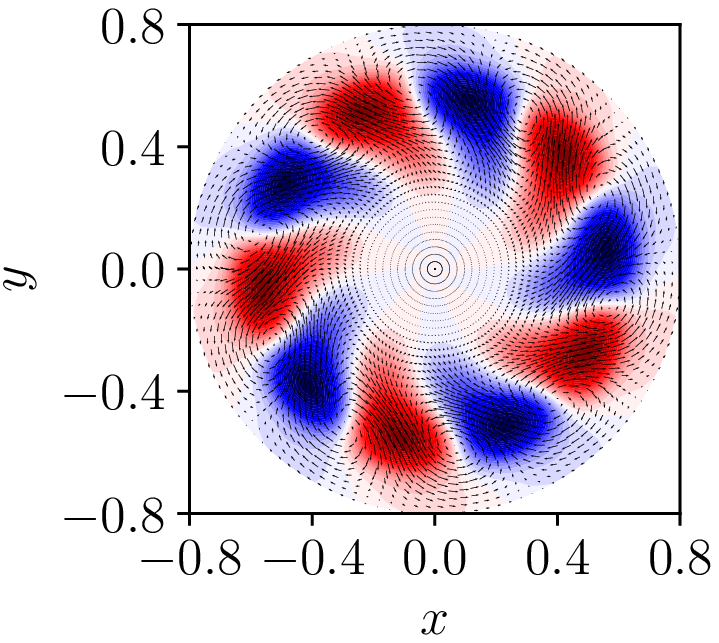}
  \label{fig:eigenFunc_excited_highSt}
  
}
\subfloat{%
\raisebox{3ex}{
  \includegraphics[scale=0.65]{figsFeb/colorbar_eigenfuntions.eps}}
  
}
\caption{$n=1$ SPOD mode corresponding to $m=5$, $St=0.03$ a) unforced b) forced at 2.6\%, $St_e=0.49$; $u_z$ are shown by contours and $u_r$, $u_\theta$ as vectors.}
\label{fig:eigenFunc_excited}
\end{figure}

The shapes of these modes are shown in figure \ref{fig:eigenFunc_excited}. At $St=0.03$ in figure \ref{fig:eigenFunc_excited_lowSt}, the $u'_z$ contours show that the structure of the streaks remains almost the same as in the unforced jet, except that they are located further away from the jet core. This could be because of the overall increased spreading of the jet with excitation. On the other hand, at $St=0.49$ in figure \ref{fig:eigenFunc_excited_highSt}, the azimuthal distortion of the streamwise vortices and streaks is seen to reduce with excitation. A further increase in the excitation level from 2.6\% to 4.8\% was seen to concentrate the energy at $St=0.03$ in fewer $m$ modes and an increase in the overall energy at $St=0.03$. 

 \section{Conclusion}
 The co-existence of high speed streaks, recently reported in \cite{nogueira2019large,pickering2020},  with other structures in turbulent round jets is studied in the current work. The energy spectra and the shape of the most energetic modes extracted from SPOD suggest a convecting system of vortex rings and streamwise vortices, which feed to the stationary streaks residing in the outer-edge of the shear layer. On strengthening the vortex rings, the shape and energies of the streaks remain almost the same. Hence, presence of a strong $m=0$ mode in turbulent jets unlike in turbulent boundary layers does not have significant effect on the occurrence of streaks.  
Having characterized these coherent structures in terms of SPOD modes, a reduced order model can be built based on these modes to study the dynamics in the near field. As our analysis here is restricted to only one downstream position, a global picture can be gained by looking at other locations.

According to the authors, more research is required in order to find the origin of streamwise vortices convecting with the vortex rings, in round jets with tripped boundary layers. Several sources of steady streamwise vortices have been proposed, such as the results of resolvant analysis by \cite{pickering2020}, where the classical lift up mechanism can give rise to steady streaks and streamwise vortices. Another possibility is that part of a deformed vortex ring could get tilted in the streamwise direction and transported into the outer region which then gives rise to steady streamwise vortices observed in the excited jets in the current study. However this needs further confirmation.

\backsection[Acknowledgements]{The authors are grateful to G. Losfeld, C. Goudeau and C. Illoul for their aid in performing HS-SPIV and J. M. Luyssen, P. Geffroy and J. P. Tobeli for their contributions towards setting up experiments.}


\backsection[Declaration of Interests]{The authors report no conflict of interest.}





\bibliographystyle{jfm}

\bibliography{references_coherentStructures_turbulentFlows,references_diffFrequencies_jet,references2020}

\end{document}